\documentclass[]{emulateapj}
\usepackage{color}

\def\ltsima{$\; \buildrel < \over \sim \;$}
\def\simlt{\lower.5ex\hbox{\ltsima}}
\def\gtsima{$\; \buildrel > \over \sim \;$}
\def\simgt{\lower.5ex\hbox{\gtsima}}

\newcommand{\cmsq}{cm$^{-2}$}
\newcommand{\nh}{$N_{\rm H}$}
\newcommand{\nhrefl}{$N_{\rm H}^{\rm refl}$}
\newcommand{\nhg}{$N_{\rm H}^{\rm Gal}$}
\newcommand{\cosmo}{($H_0$, $\Omega_{\rm m}$, $\Omega_{\lambda}$)}

\shorttitle{{\it SUZAKU} VIEW OF THE {\it SWIFT}/BAT ACTIVE GALACTIC NUCLEI. I\hspace{-.1em}V.}
\shortauthors{Tazaki et al.}

\begin{document}

\title{{\it SUZAKU} VIEW OF THE {\it SWIFT}/BAT ACTIVE GALACTIC NUCLEI. I\hspace{-.1em}V. \\
NATURE OF TWO NARROW-LINE RADIO GALAXIES (3C 403 AND IC 5063)}
\author{
 Fumie Tazaki\altaffilmark{1},
 Yoshihiro Ueda\altaffilmark{1},
 Yuichi Terashima\altaffilmark{2},
 and Richard F. Mushotzky\altaffilmark{3}
}

\altaffiltext{1}{Department of Astronomy, Kyoto University, Kyoto 606-8502, Japan}
\altaffiltext{2}{Department of Physics, Ehime University, Matsuyama 790-8577, Japan}

\altaffiltext{3}{Department of Astronomy, University of Maryland, College Park, MD, USA}

\begin{abstract} 

We report the results of {\it Suzaku} broad band X-ray observations of
the two narrow-line radio galaxies (NLRGs), 3C 403 and IC 5063.
Combined with the {\it Swift}/BAT spectra averaged for 58 months, we are
able to accurately constrain their spectral properties over the 0.5--200
keV band. The spectra of both nuclei are well represented with an
absorbed cut-off power law, an absorbed reflection component from cold
matter with an iron-K emission line, and an unabsorbed soft component,
which gives a firm upper limit for the scattered emission. The
reflection strength normalized to the averaged BAT flux is $R \equiv
\Omega/2\pi \approx 0.6$ in both targets, implying that their tori have
a sufficiently large solid angle to produce the reprocessed
emission. A numerical torus model with an opening angle of $\sim
50^\circ$ well reproduces the observed spectra. We discuss the
possibility that the amount of the normal gas responsible for Thomson
scattering is systematically smaller in radio galaxies compared with
Seyfert galaxies.

\end{abstract}


\section{Introduction}

Modern astronomy has revealed that the mass of a super-massive black
holes (SMBHs) in a galactic center has tight correlation with the mass
of the bulge of the host galaxy \citep[e.g.,][]{Mag98, Mar03},
indicating the co-evolution between SMBHs and galaxies. According to
theoretical predictions, active galactic nuclei (AGNs) may play a key
role in regulating both mass accretion and star formation by their
feedback through powerful outflows and/or radiation to the surrounding
gas \citep[e.g.,][]{DiM05}. Broadband X-ray observations of AGNs are
particularly important to study the physics of accretion and outflow in
the central engine by measuring the spectra and time variability of
the continuum emission from the innermost disk and relativistic
jets. They also provide unique information on the structure of the tori
surrounding the accretion disk, which may be the source of mass
accretion and site of nuclear starbursts
\citep[e.g.,][]{Dav07,Ima04}, through investigation of its
reprocessed emission.

Recent hard X-ray all sky surveys above 10 keV conducted by {\it
Swift}/BAT and {\it INTEGRAL} are providing us with an ideal sample of
AGNs in the local universe that can be efficiently followed-up with
observatories having a narrow field-of-view, like {\it Suzaku} and
{\it XMM-Newton}, for detailed spectroscopic studies. {\it Suzaku} has
unique features covering the wide energy band over 0.1--600 keV with
good energy resolution and large collecting area below 10 keV. Making
use of these capabilities, we have been working on a program of {\it
Suzaku} follow-up observations of {\it Swift}/BAT AGNs, mostly
focusing on heavily obscured ones \citep{Egu09,Win09,Egu11}. A
remarkable finding is the discovery of ``new type'' AGNs that exhibit
a very small amount of scattering gas, which are interpreted to have
geometrically very thick tori, unlike classical type AGNs
\citep{Ued07,Egu09}.

Radio galaxies, a class of AGNs exhibiting powerful jets, are key
objects to understand the origin and effects of the AGN feedback to
the surroundings. However, the fundamental question, how the structure
of the nucleus is different between AGNs with jets and without jets, is not fully resolved yet. This paper is the fourth in a series related to {\it Suzaku} and {\it Swift}/BAT observations of AGN, targeting two narrow line radio galaxies (type-2 radio loud AGNs) 3C~403 and IC~5063, for which no
``simultaneous'' broad band spectra over the 0.1--200 keV band have
been reported. All targets studied in our previous series of papers
are Seyfert galaxies (i.e., AGNs without powerful jets) except for
NGC~612 reported in \citet{Egu11}. By adding these new targets, 
we increase the sample for comparing the statistical properties 
of radio loud AGNs to radio quiet ones.

\begin{deluxetable*}{ccc}
\tabletypesize{\footnotesize}
\tablecaption{List of Targets\label{targets}}
\tablewidth{0pt}
\tablehead{\colhead{Target Name ({\it Swift} ID)} & \colhead{3C 403 (J1952.4+0237)} & \colhead{IC 5063 (J2052.0--5704)}}
\startdata
R.A. (J2000)\tablenotemark{a} & 19 52 15.809 & 20 52 02.34 \\
Decl. (J2000)\tablenotemark{a} & +02 30 24.18 & --57 04 07.6 \\
Redshift\tablenotemark{a} & 0.059 & 0.011348 \\
{\it Suzaku} Observation ID & 704011010 & 704010010\\
Start Time (UT) & 2009-04-08T19:53:00 & 2009-04-24T15:04:25 \\
End Time (UT) & 2009-04-10T02:35:24 & 2009-04-25T16:28:19 \\
Exposure\tablenotemark{b} (XIS) (ks) & 47.9 & 45.2 \\
Exposure (HXD/PIN) (ks) & 39.3 & 57.6
\enddata
\tablenotetext{a}{The position and redshift for each source are taken from the NASA/IPAC Extragalactic Database.}
\tablenotetext{b}{Based on the good time interval of XIS-0.}
\end{deluxetable*}

3C 403 ($z$ = 0.059) is a NLRG classified as ``X-shaped'' or
``winged'' radio morphology \citep{Lea84}
, which is explained as the results of propagation 
of jets in non-axisymmetric atmospheres \citep{Kra05}.
The host galaxy Hubble type is S0 or peculiar.
\citet{Vas10} obtained the mass of the SMBH to be $10^{8.26}$
$M_{\odot}$ using the correlation between black hole mass and bulge
luminosity, according to the procedure described in \citep{Vas09}.
The bulge luminosity is derived from the total K-band flux in the
2MASS point source catalog by subtracting the AGN contribution estimated
from the X-ray luminosity.
\citet{Kra05} studied the X-ray properties of 3C 403 with {\it
Chandra}, detecting X-ray emission from the nucleus, hot interstellar
medium (ISM), extended radio structures (lobes and wings), and compact
regions in the jet (knots and hot spots).  IC 5063 ($z$ = 0.0110) has
a highly anisotropic ionizing radiation field with ``X'' or
``conical'' morphology \citep{Col91}, associated with a giant
elliptical or S0 host galaxy. The black hole mass is estimated to be
$10^{7.41}$ $M_{\odot}$ \citep{Vas10}. \citet{Koy92} observed this
object with the {\it Ginga} satellite and found that the energy
spectrum in the 3--18 keV band is described by a power-law with a
photon index of 1.5 and an absorption column density of $2 \times
10^{23}$ \cmsq. \citet{Vig97} report the results of observations with
{\it ASCA} and {\it ROSAT}. From the 0.1--10 keV spectrum of IC 5063,
they detect two power-law components (with photon indices of $
1.7\pm0.2$ and $2.2\pm0.3$, where the former is absorbed with \nh $\sim 2
\times 10^{23}$ \cmsq), and a narrow iron-K$\alpha$ line at $E_{\rm
K\alpha} \approx$ 6.4 keV (with an equivalent width of $\sim$200 eV).

The organization of this paper is as follows. Section~2 summarizes the
observations. Data analysis and results are presented in Section~3. We
discuss the implication of our results in Section~4. The cosmological
parameters \cosmo = (70 km s$^{-1}$ Mpc$^{-1}$, 0.3, 0.7;
\citealt{Kom09}) are adopted in calculating the luminosities. The
errors attached to spectral parameters correspond to those at 90\%
confidence limits for a single parameter.

\section{Observations}\label{obs}

{\it Suzaku}, the fifth Japanese X-ray satellite \citep{Mit07},
carries a set of X-ray CCD cameras called the X-ray Imaging Spectrometer (XIS)
and a non-imaging instrument called the Hard X-ray Detector (HXD)
composed of Si PIN photodiodes and Gadolinium Silicon Oxide (GSO)
scintillation counters. In this paper, we analyze the data of three XISs and HXD/PIN, which cover the energy band of 0.2--10 keV
and 10--60 keV, respectively. The data of HXD/GSO covering energies above 50
keV are not utilized because the fluxes of our targets are too faint
to be detected. In the spectral analysis, we also use the spectra of
{\it Swift}/BAT in the 14--195 keV band averaged for 58-months
\citep{Bau11}.

The two NLRGs 3C 403 and IC 5063, cataloged in the {\it Swift}/BAT
9-months survey \citep{Tue08}, were observed with {\it Suzaku} in 2009
April. The basic
information of our targets and observation log are summarized in
Table~\ref{targets}. Each object was observed at the ``HXD nominal''
pointing position for a net exposure of $\sim$50 ks after data
screening.

\section{Analysis and Results}\label{ana}

The data reduction was performed according to standard procedures from
the cleaned event files provided by the {\it Suzaku} team (processing
version 2.3.12.25). We used FTOOLS (heasoft version 6.8) for
extraction of light curves and spectra, and XSPEC (version 12.5.1n)
for spectral fitting. The XIS events of the source were extracted from
circular regions centered on the source peak with a radius of 2'
(3C~403) and 2'.5 (IC~5063), 
in which 85\% and 91\% of the total source photons are accumulated
by the X-ray telescopes, respectively.
The background data were taken from annulus regions centered at the
averaged optical axis of the X-ray telescopes with radii between
2.8'--6.7' for 3C 403 and 1.2'--6.3' for IC 5063, by excluding regions
where the target and other bright sources are located.
We generated the RMF files of the XIS with {\it xisrmfgen}, and the ARF 
files with {\it xissimarfgen} \citep{Ish07}. The ``tuned'' NXB event 
files provided by the HXD team were utilized to produce the background 
spectra, to which that of the cosmic X-ray background (CXB) was added 
based on the formula of \citet{Gru99}. 
The source flux of HXD/PIN in the 15--40 keV band corresponds to
$\simeq 6$\% (3C~403) and $\simeq 24$\% (IC~5063) of the background
(NXB+CXB) rate. 
The signal-to-noise ratio calculated as the source photon counts
divided by the square root of the total (source plus background)
counts  in this band is 5.3 for 3C~403 and 26.7 for IC~5063.
The statistical error is larger than the
systematic uncertainty of $\simeq 0.34\%$ (1$\sigma$) in the
background model for a 40 ks exposure \citep{Fuk09}.  We used
ae\_hxd\_pinhxnome5\_20080716.rsp for the HXD/PIN response.

\subsection{Light Curves}\label{LC}

\begin{figure}
\begin{center}
\epsscale{0.85}
\rotatebox{-90}{
\plotone{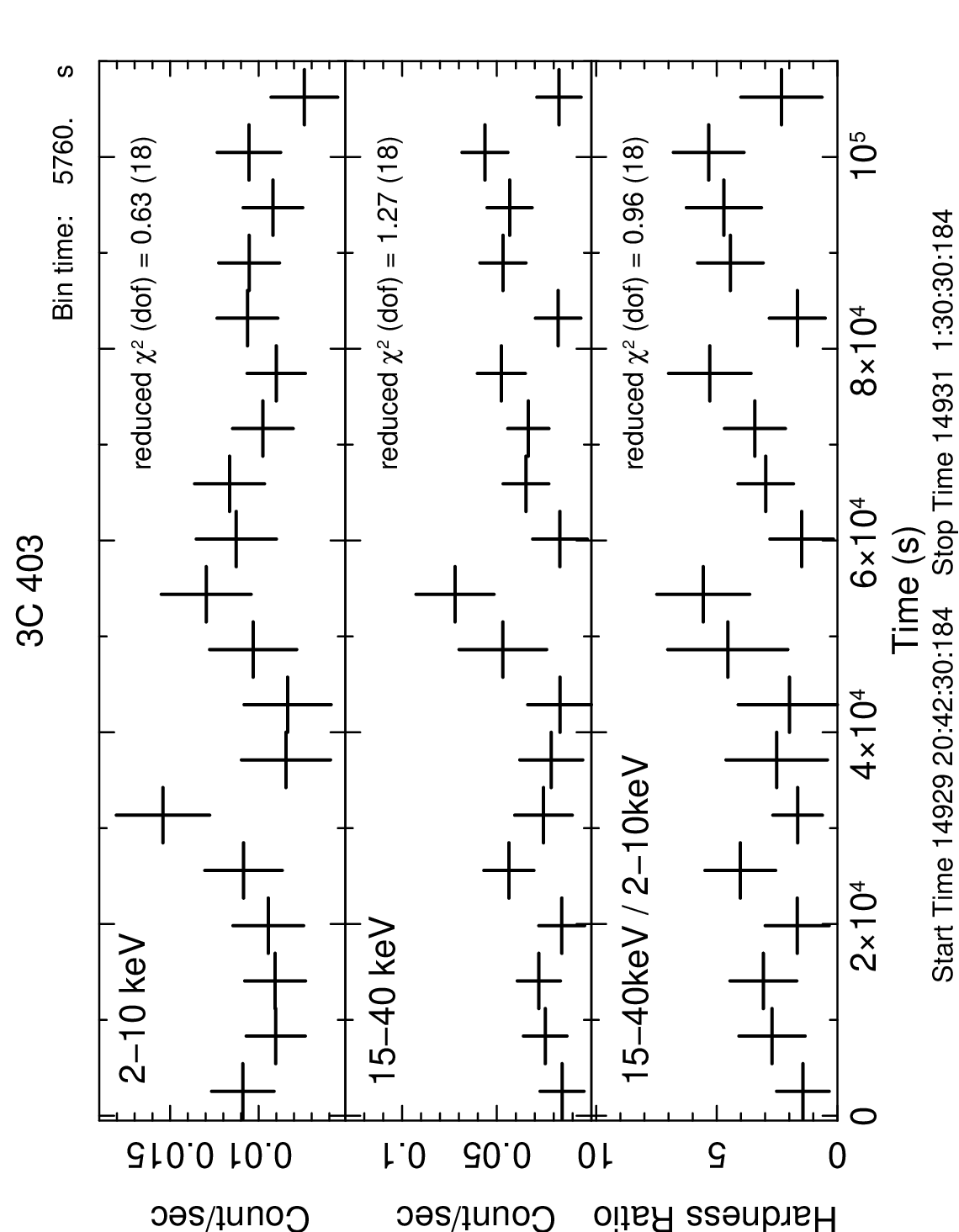}
\plotone{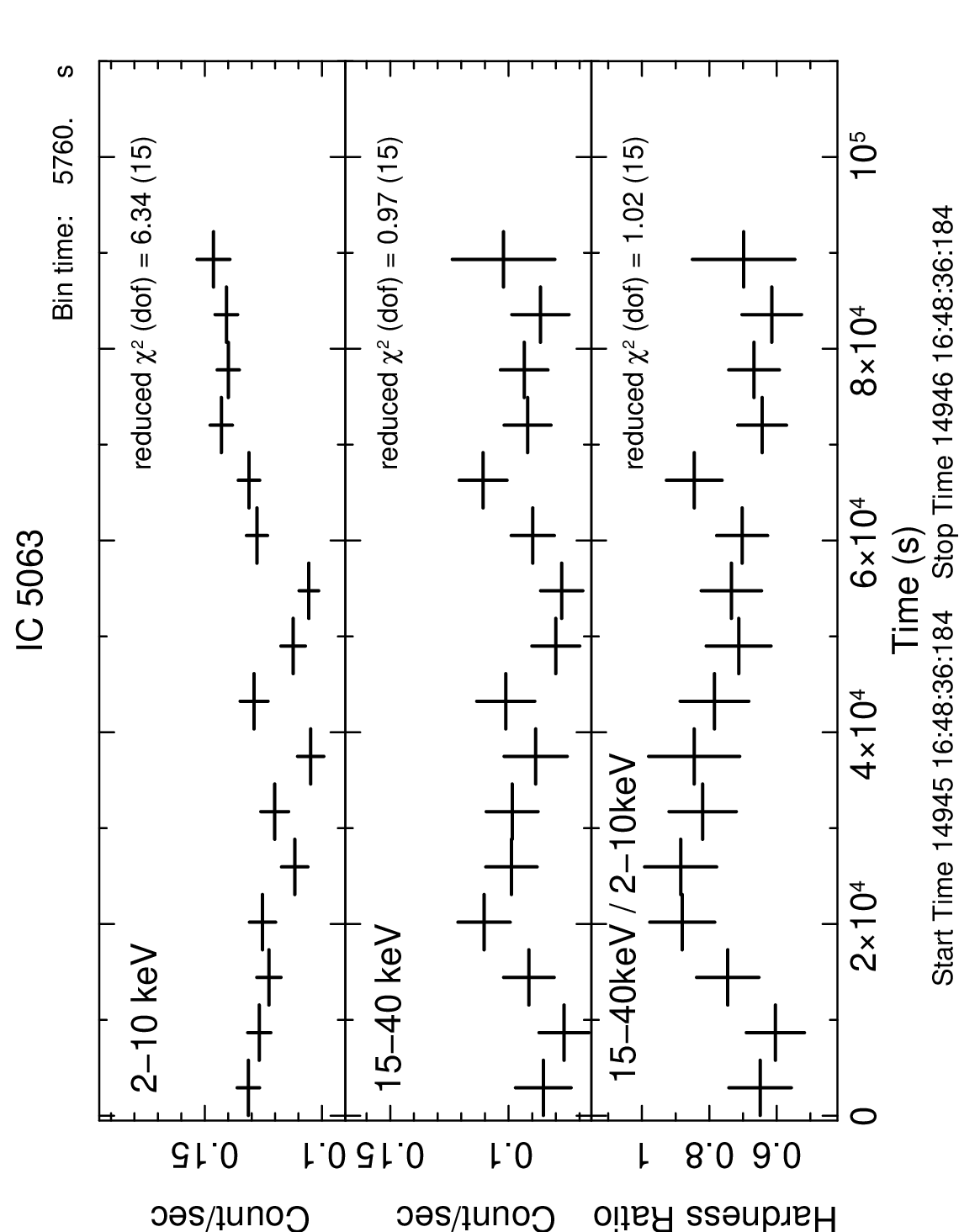}
}
\caption{ 
({\it top}) Light curves of 3C~403 obtained with the XIS in the 2--10 keV band (upper) and with the HXD/PIN in the 15--40 keV band (middle), and the hardness ratio between the two bands (lower). The bin size is 5760 s. The reduced $\chi^2$ with the degrees of freedom for the constant flux hypothesis is shown at the upper right corner in each panel.   
({\it bottom}) The same as IC~5063.
}
\label{Suzaku_LC}
\end{center}
\end{figure}

Figure~\ref{Suzaku_LC} shows the light curves of 3C 403 and IC 5063 in
the 2--10 keV band combined from XIS-0 and XIS-3 (upper), those in the
15--40 keV band from HXD/PIN (middle), and their hardness ratio
between the two bands. 
A bin width of 5760 s, the
orbital period of {\it Suzaku} (including periods of data gap
that are excluded to calculate the count rate), is chosen
to eliminate any systematics caused by the orbital change of 
the satellite. There is no evident time
variability in the light curves of 3C 403, while the 2--10 keV flux of
IC 5063 increases after $\sim 6 \times 10^4$ s from the beginning of
the observation. The constant flux model is rejected by with reduced
$\chi^2$ of 6.34 with 15 degrees of freedom. Due to
the limited statistics in the PIN data, however, the 15--40 keV light
curve and hardness ratio of IC~5063 show no significant variability
based on $\chi^2$ tests. We examine the spectral variability of
IC~5063 in Section~\ref{IC 5063}.

\subsection{BAT Spectra}\label{BAT_spec}

Before performing spectral fitting to the {\it Suzaku} data, we analyze
the {\it Swift}/BAT spectra in the 14--195 keV band to constrain the
high energy cutoff. The detection significance of 3C~403 and
IC~5063 in the 58-months data is 9.6 and 26.8, respectively (Baumgartner
et al.\ 2011).  A simple power law fit with $\propto E^{-\Gamma}$ is
statistically acceptable, yielding a photon index of $\Gamma = 1.7\pm0.3$
with $\chi^2/{\rm dof}=8.5/6$ for 3C~403, and $\Gamma = 1.9\pm0.1$ with
$\chi^2/{\rm dof}=3.7/6$ for IC~5063. It is known, however, that the
hard X-ray continuum of AGNs is generally described by a power law with
an exponential cutoff, $\propto E^{-\Gamma} {\rm exp}(-E/E_{\rm cut})$,
where $E_{\rm cut}$ is a cutoff energy. Since the cutoff energy is
typically $> 100$ keV \citep[e.g.,][]{Del03}, the {\it Suzaku}
data alone (XIS and HXD/PIN) cannot determine it.  To fit the BAT
spectra, we adopt {\bf pexrav} model \citep{Mag95} in XSPEC, which
contains a Compton reflection component from optically thick cold
matter. The strength of reflection is represented with $R \equiv
\Omega/2\pi$, where $\Omega$ is the solid angle of the reflector viewed
from the irradiating source. In this analysis, we assume two extreme
cases for the reflection strength, $R$ = 0 and 2. The inclination
angle in the {\bf pexrav} model is fixed at $60^{\circ}$ as a
representative value throughout our paper except in Section~\ref{torus_model}. 
To avoid strong coupling
between $E_{\rm cut}$ and $\Gamma$, we adopt $\Gamma=1.7$ for both
targets, which is an averaged value of 13 nearby radio galaxies
obtained from previous X-ray studies \citep{Gra06}. As summarized in
Table~\ref{BAT_spec}, we find that $E_{\rm cut}$ $>$
120 keV for 3C~403 and $E_{\rm cut}$ = 130--470 keV for
IC~5063. In the following spectral analysis, we fix the cutoff energy
at 200 keV for both 3C 403 and IC 5063, since
changing these values within the above uncertainties does not affect
our conclusions.

\begin{deluxetable}{cccc}
\tabletypesize{\footnotesize}
\tablecaption{Cutoff Energies ($E_{\rm cut}$) Estimated by the BAT spectra\label{BAT_spec}}
\tablewidth{0pt}
\tablehead{\colhead{} & \colhead{Target Name} & \colhead{3C 403} & \colhead{IC 5063}}
\startdata
$R = 0$ & $E_{\rm cut}$ & $540 (> 120)$ & $220^{+250}_{-80}$ \\
 & $\chi^2$/dof & 8.9/6 & 3.3/6 \\ \hline
$R = 2$ & $E_{\rm cut}$ & $570 (> 140)$ & $190^{+140}_{-60}$ \\
 & $\chi^2$/dof & 8.6/6 & 12.9/6
\enddata
\tablecomments{The photon index is fixed at 1.7 for both targets.}
\end{deluxetable}

\subsection{Spectral Analysis with Analytic Models}\label{spec_model}

\begin{figure}
\begin{center}
\epsscale{0.9}
\rotatebox{-90}{
\plotone{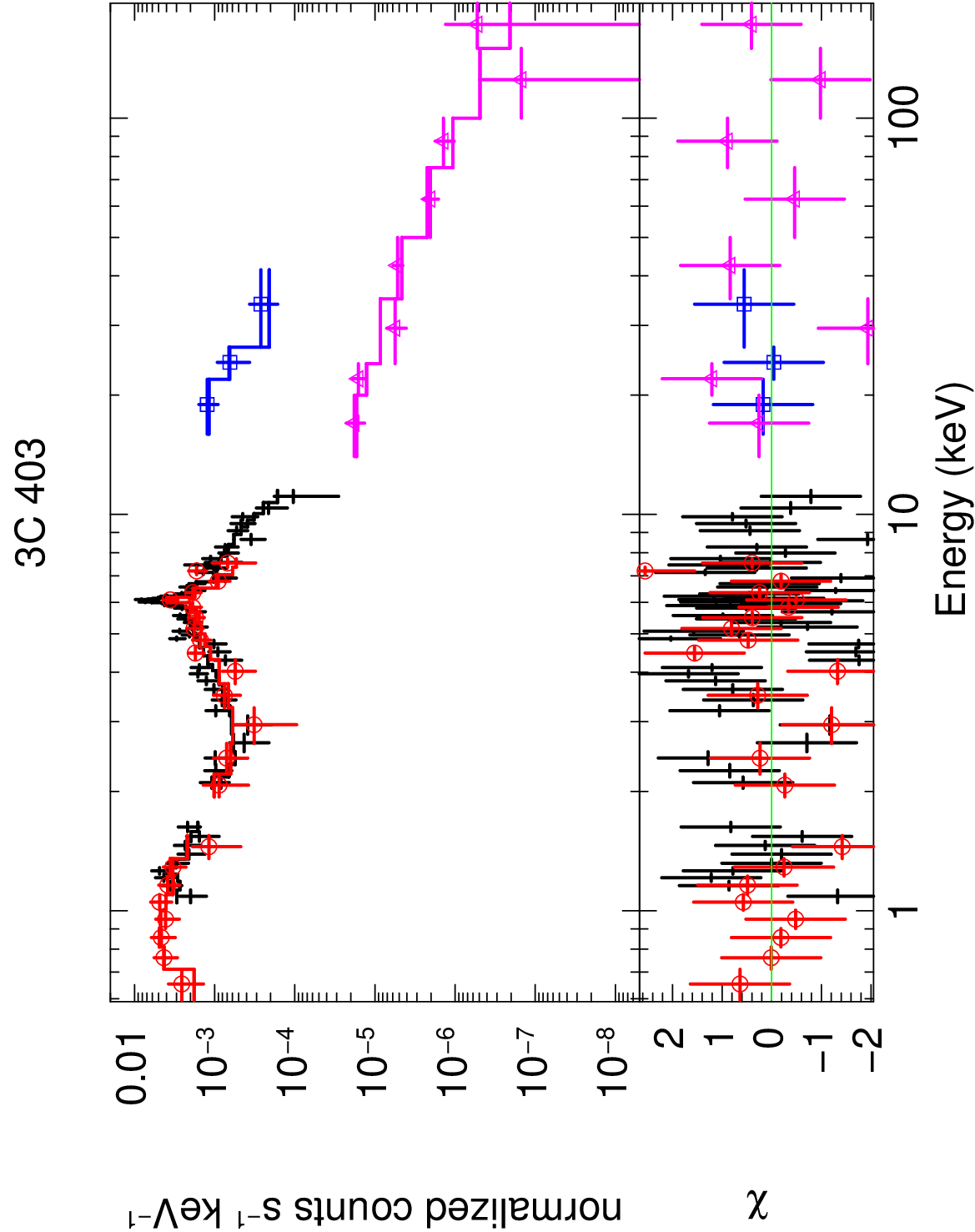}
\plotone{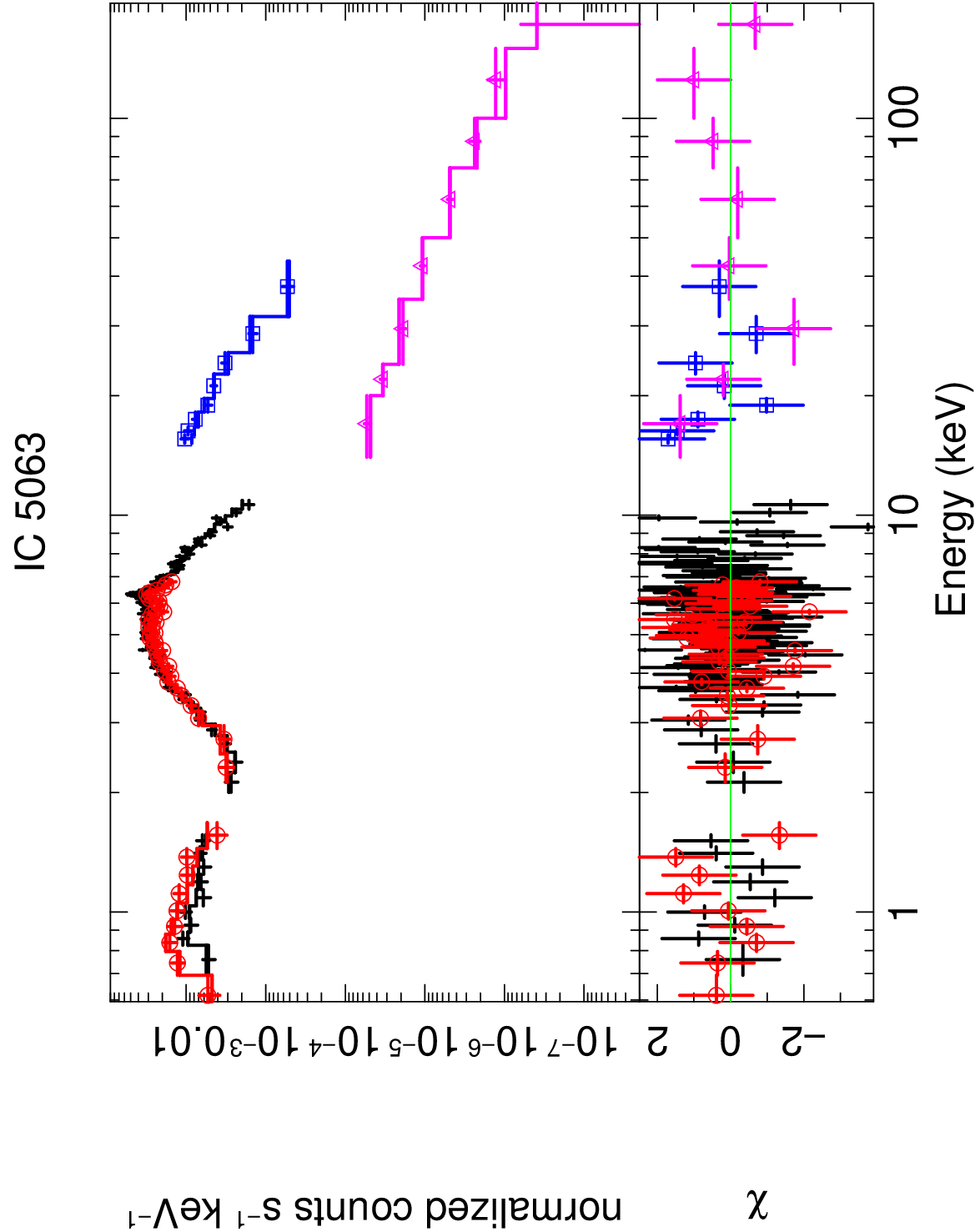}
}
\caption{ 
({\it top}) The folded spectra of 3C~403 obtained with the XIS-FIs in the 1--11 keV band (black), the XIS-BI in the 0.5--8 keV band (red, open circles), the HXD/PIN in the 16--40 keV band (blue, open squares), and the {\it Swift}/BAT in the 14--195 keV band (magenta, open triangles). The best-fit model (B in Table \ref{para}) is plotted by the solid lines, and the residuals in units of $\chi$ are shown in the lower panels. 
({\it bottom}) The same as IC 5063.
}
\label{data}
\end{center}
\end{figure}

For spectral analysis, we use the {\it Suzaku} spectra of the 2 XIS-FIs 
in the 1--11 keV band, and the XIS-BI in the 0.5--8 keV band, 
and HXD/PIN in the 16--40 keV band, where the highest signal-to-noise ratios are
achieved. The XIS data in the 1.7--1.9 keV band are discarded because
of calibration uncertainties associated with the instrumental Si-K
edge. The relative flux normalization of the XIS-BI to the XIS-FIs is
set free, while that of the HXD/PIN to the XIS-FIs is fixed at 1.18
based on the calibration using the Crab Nebula \citep{Mae08}. The
Galactic absorption, \nhg, is included in the spectral model, which is
fixed at the value derived from the H~I map of \citet{Kal05} (\nhg =
$1.22 \times 10^{21}$ \cmsq for 3C~403 and $6.10 \times 10^{20}$ \cmsq
for IC~5063). Solar abundances by \citet{And89} are assumed throughout
our analysis. The upper panels of Figure~\ref{data} plot the {\it Suzaku} spectra of the two sources folded with the detector responses,
together with the BAT spectra in units of photons cm$^{-2}$ s$^{-1}$.

We apply the three analytic models defined in \citet{Egu09}, Models~A,
B, and C (see below), to which we add a spectral component from an
optically thin plasma with Solar abundances ({\bf apec}) because it
improves the fit significantly in both targets. In the case of 3C~403,
two power law components from the radio lobes/wings and from the
compact structures in the jets are also included. 
These $\sim$10" scale structures are spatially resolved in
the {\it Chandra} data \citep{Kra05};
their fluxes in the 0.25--10 keV band are estimated to be
$(4.4\pm0.3)\times10^{-14}$ erg cm$^{-2}$ s$^{-1}$ and
$(3.5\pm0.3)\times10^{-14}$ erg cm$^{-2}$ s$^{-1}$, corresponding to
0.31\% and 0.27\% of the total flux from the unresolved nucleus
corrected for absorption, respectively.

Model A consists of a direct component from the nucleus (a cutoff
power law absorbed by cold matter), a scattered component (a cutoff
power law without absorption), and an iron-K emission line (a
Gaussian). The high energy cutoff is fixed at 200 keV based on the BAT
only results (see above). 
The scattered component is assumed to have the same slope as the
direct one with a fraction of $f_{\rm scat}$. In Model B, a
Compton reflection component of the direct continuum from optically thick
matter ({\bf pexrav} model), absorbed with a column density of
$N_{\rm H}^{\rm refl}$, is further added to Model A. This emission is
expected from the inner wall of the torus and/or the accretion disk
irradiated by the central source. The absorption for the reflection
component is set to be independent of that of the transmitted one,
because the emission region is different and the line-of-sight column
density may not always be the same, depending on the viewing
geometry. Model C has the same emission components as Model B
including an absorbed reflection component, but we consider two layers
of absorber with different hydrogen column densities for the
transmitted component, corresponding to the ``partial covering'' case
where the absorbers in the line of sight consist of gas blobs smaller
than the size of the X-ray emitting region \citep[e.g.,][]{Eli06}.

We first perform simultaneous fit only to the {\it Suzaku} XIS and
HXD/PIN spectra to select the most appropriate model for each object
among the three. To avoid complex effects of time variability,
the BAT spectra averaged for 58 months are not utilized in this stage. 
We start from the simplest model (Model~A), and adopt a
more complex model (Models B or C) only if we find a statistically
significant improvement of the fit or a physically more reasonable
solution. After selecting the best model describing the data in this
way, we finally include the BAT spectra in the joint fit to better
constrain the continuum up to 200 keV. We allow only the
normalization of the direct component to vary while keeping the
continuum shape (i.e., spectral slope and cutoff energy) the same
between the {\it Suzaku} and {\it Swift}/BAT epoch. The normalizations
of the reflection and scattered component are also linked between {\it
Suzaku} and {\it Swift}/BAT, considering that the time scale of their
variability should be larger than $>$years if the emission region size
is typical ($>$1 pc) of the scale size of tori. Unlike our previous papers
\citep{Ued07,Egu09,Egu11}, we define the reflection strength and
scattered fraction relative to the BAT flux, not to the {\it Suzaku}
flux, since the BAT spectra integrated over 58 months should be a good
indicator of the averaged flux level responsible for these reprocessed
emission.

\begin{figure}
\begin{center}
\epsscale{0.9}
\rotatebox{-90}{
\plotone{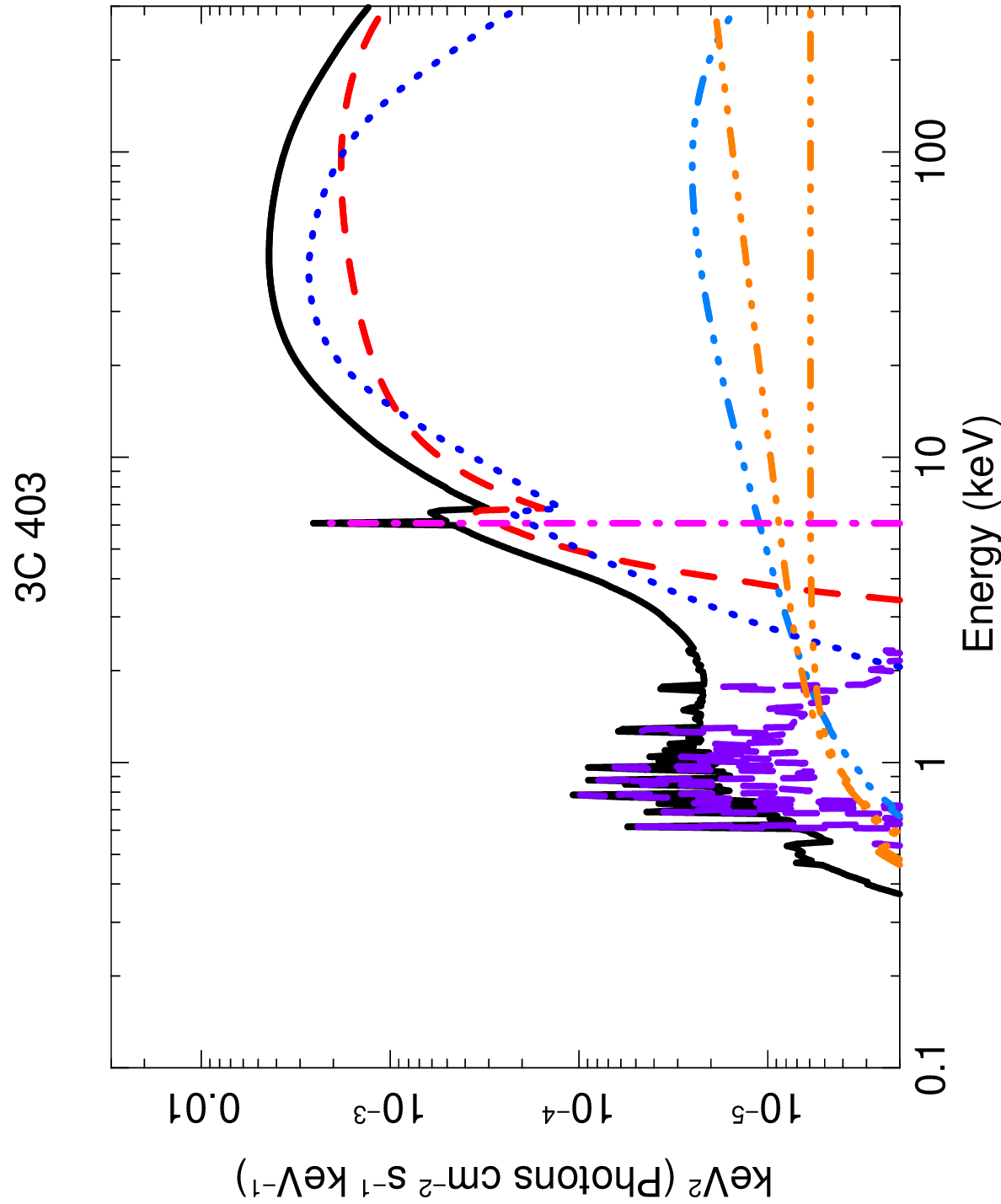}
\plotone{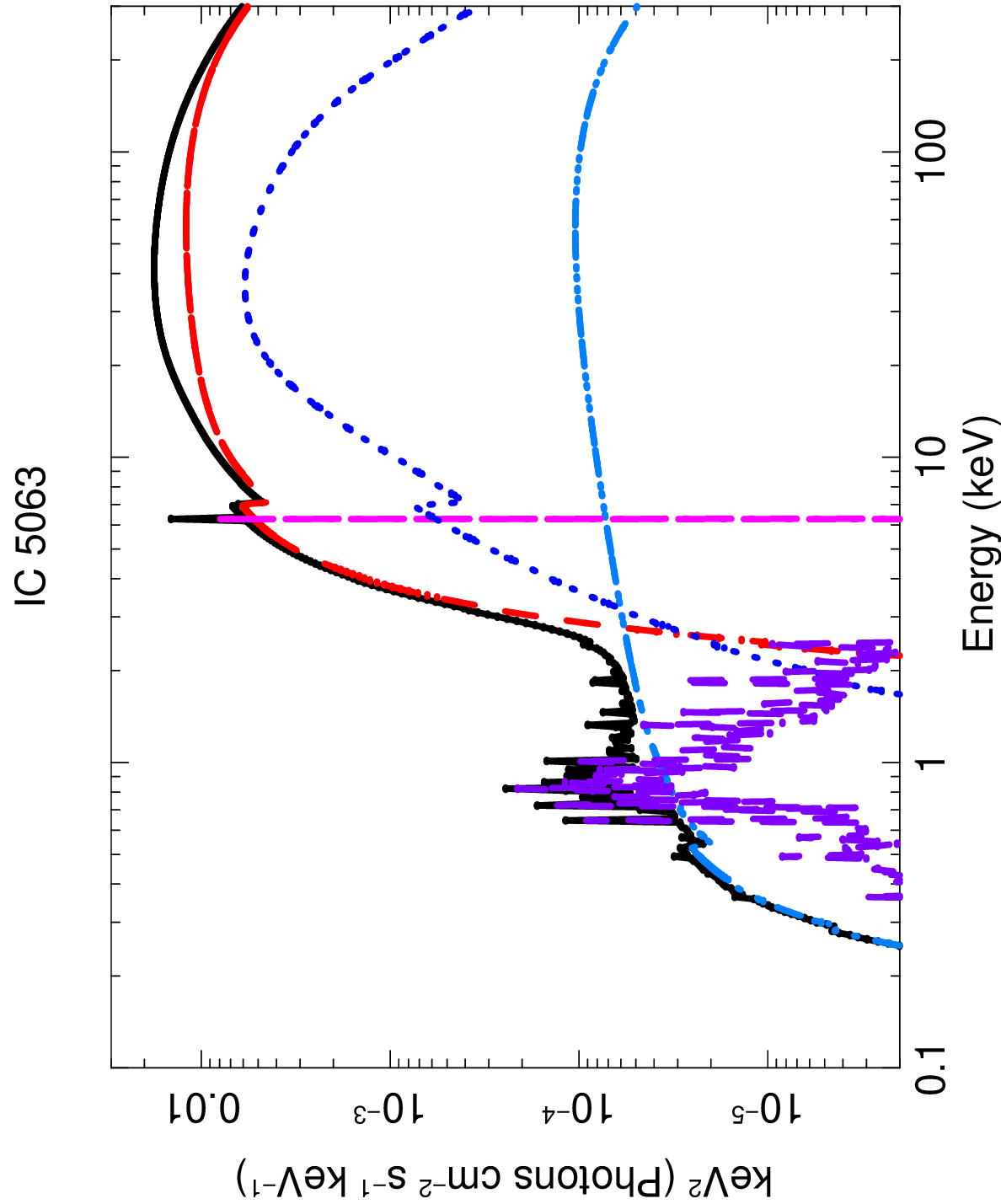}
} 5.4754

\caption{ 
({\it top}) Best-fit analytic model for 3C~403 in units of $E F(E)$
($F(E)$ is the energy flux at the energy $E$). The solid (black), dashed (red),
dotted (blue), and dot-dashed (magenta) curves correspond to the total,
direct component, reflection component, and iron-K emission line. The
dashed (purple) curve represents the emission from an optically thin
thermal plasma, and dot-dot-dot-dashed (orange) curves show the emission
from the jets and radio lobes/wings (see the text). The dot-dot-dot-dashed (cyan) curve represents the scattered component. 
({\it bottom}) The same as IC~5063. 
}
\label{model}
\end{center}
\end{figure}

\subsubsection{3C 403}\label{3C 403}

From the results by \citet{Kra05} obtained with {\it Chandra}, we are
able to estimate the X-ray fluxes from the radio lobes/wings and
compact features in the jets (knots and hot spots) that should
contribute to the {\it Suzaku} spectra, where these jet-related
structures and the nucleus are not spatially resolved.
Since neither of these components are expected to be variable 
on time scale of years, we fix their fluxes at the best-fit {\it Chandra} 
values; we confirm that the uncertainties do not affect our results as their
contributions in the total flux are minor.
The component from the radio lobes/wings and jets are represented as a
power law with a photon index of 1.8 and 2.0 with a normalization at 1
keV of $9.8 \times 10^{-15}$ and $9.6 \times 10^{-15} {\rm erg \
cm^{-2} s^{-1} keV^{-1}}$, respectively, both are always added to the
above three models representing the nucleus emission. 
Since photon index of the power law in the lobes/wings component is 
poorly constrained from the {\it Chandra} data, we fix its slope to that of the radio emission \citet{Den02}, assuming that the radio and X-ray emission 
correspond to the synchrotron and inverse Comptonization from the same electrons,
respectively. To model optically thin thermal emission from the hot
interstellar medium (ISM) detected by \citet{Kra05}, we also add an
{\bf apec} model whose temperature and emission measure are free 
parameters. The elemental abundances are fixed at the Solar values.

We adopt Model~B as the most appropriate model for the nucleus
emission of 3C~403. Model A yields a very small photon index of the
transmitted component, $\sim$1.0, which seems rather
unphysical, with $\chi^2/{\rm dof}$ = 82/73.
Reasonable results are obtained with Model~B, which gives
a photon index of $\sim$1.5 with $\chi^2/{\rm dof}$ = 74/71. Model~C
does not improve the fit significantly in terms of the $\chi^2$ value
($\chi^2/{\rm dof}$ = 74/69). The best fit model (Model~B + two power
laws + apec) to the joint fit including the {\it Swift}/BAT
spectrum is overplotted in the upper panel of Figure~\ref{data} with
residuals in the lower panel. Figure~\ref{model} (a) plots its $E
F(E)$ form, where $F(E)$ is the energy flux at the energy $E$. 

The best-fit parameters are summarized in Table~\ref{para}.  The
moderately flat photon index ($\Gamma =1.53^{+0.03}_{-0.02}$), 
though within the typical range for radio galaxies \citep{Gra06},
may be affected by simplification of the analytic model where only 
a single absorber is assumed for the reflection component. 
The scattered fraction $f_{\rm
scat} \sim 0.4\%$ is rather small, corresponding to that of a ``new
type'' AGN in \citet{Ued07} if it corresponds to the opening angle of
the torus. This issue will be discussed in Section~4. The iron-K
emission line is detected at $\approx 6.4$ keV and is narrow (with a
$1\sigma$ width of $<$98 eV at 90\% confidence level). This indicates
that it mainly comes from cold and distant matter from the black hole,
fully consistent with a torus origin. The equivalent width of the
iron-K line with respect to the reflection continuum is 1.13$\pm$0.24
keV, which is physically self-consistent according to theoretical
calculations \citep[e.g.,][]{Ike09}. The smaller column density for the reflected
component ($N_{\rm H}^{\rm refl} \sim 4\times10^{22}$ cm$^{-2}$) than
that for the transmitted one (\nh $\approx 6 \times 10^{23}$ \cmsq)
suggests that we are looking at the source with a somewhat face-on
geometry so that a significant fraction of the reflected light from
the inner wall of the torus can reach us without strong obscuration.

While we detect a very strong reflection within the {\it Suzaku}
spectra corresponding to $R \approx 2$, which dominates the HXD/PIN
flux, we find a lower value of $R\approx 0.6$ when normalized to the
flux of the transmitted component determined by {\it Swift}/BAT. This
can be explained by time variability since the flux during the {\it
Suzaku} observation was about 3 times smaller than the averaged flux
over the 58 months of the BAT observations. In fact, the difference
between the unabsorbed nuclear flux in the 0.25--10 keV band was $1.3
\times 10^{-11}$ erg cm$^{-2}$ s$^{-1}$ at the {\it Chandra}
observation in 2002, while that measured with {\it Suzaku} is $4.0
\times 10^{-12}$ erg cm$^{-2}$ s$^{-1}$. 

The best fit temperature of the {\bf apec} component is 
$kT = 0.29\pm0.03$ keV, consistent with the {\it Chandra}
result \citep{Kra05}, $kT = 0.3\pm0.04$ keV. We note, however,
that its flux as determined with {\it Suzaku}, $2.8 \times 10^{-13}$
erg cm$^{-2}$ s$^{-1}$ in the 0.25--10 keV band, is about 5 times
higher than that estimated from the {\it Chandra} data, even taking
into account the contribution from the hot ISM in an outer region at
radii larger than 7".5 up to 60" from the nucleus (see \citet{Kra05}
Section 3.1), which is contained in the {\it Suzaku} extraction 
region. We find that this discrepancy is mainly produced due to the 
spectral coupling with the unabsorbed soft component; its slope is 
assumed to be the same as for the direct component in our spectral model, 
$\Gamma=1.5$, which is smaller than that adopted by \citet{Kra05},
$\Gamma=2.0$. Thus, the difference may be regarded as a systematic
uncertainty in the absolute flux from the hot ISM, whose core
emission is difficult to spatially resolve even with {\it Chandra}
from the AGN component. We confirm that our result on the amount of
the scattering gas presented in Section~\ref{3C 403} is not
significantly affected (within a factor of 1.5) even when we fix the
flux of the {\bf apec} component at the {\it Chandra} value in the
spectral analysis.

\begin{deluxetable*}{cccc}
\tabletypesize{\footnotesize}
\tablecaption{Best-Fit Parameters of {\it Suzaku} and {\it Swift}/BAT spectra with Analytic Model\label{para}}
\tablewidth{0pt} 
\tablehead{\colhead{} & \colhead{Target} & \colhead{3C 403} & \colhead{IC 5063}}
\startdata
     & Best-fit model & B + apec + two power laws & B + apec \\
(1)  & \nhg ($10^{22}$ \cmsq) & 0.122 & 0.061 \\
(2)  & \nh ($10^{22}$ \cmsq) & $61^{+6}_{-5}$ & $25.0 \pm 0.7$ \\
(3)  & Norm$_{\rm BAT}$ & $3.2 \pm 0.7$ & $1.06 \pm 0.08$ \\
(4)  & $\Gamma$  & $1.53^{+0.03}_{-0.02}$ & $1.72 \pm 0.02$ \\
(5)  & $f_{\rm scat}$(\%)  & $0.4 \pm 0.3 $ & $0.9 \pm 0.1$ \\
(6)  & $E_{\rm cen}$ (keV)   & $6.43^{+0.04}_{-0.02}$ & $6.39^{+0.02}_{-0.08}$ \\
(7)  & $E_{\rm wid}$ (eV)  & $<$ 98 & $<$ 78  \\
(8)  & EW (eV)    & $455 \pm 98$ & $148 \pm 20$  \\
(9)  & EW$^{\rm refl}$ (keV) & $1.13 \pm 0.24$ & $1.52 \pm 0.21$ \\ 
(10)  & \nhrefl ($10^{22}$ \cmsq)   & $3.6^{+4.2}_{-2.8}$ & $3.1^{+3.4}_{-1.9}$   \\
(11)  & $R$  & $0.65 \pm 0.17$  & $0.65 \pm 0.13$  \\
(12)  & $kT$ (keV) & $0.29 \pm 0.03$ & $0.61^{+0.07}_{-0.08}$ \\
(13)  & $n^2V$ (cm$^{-3}$) & $(2.7 \pm 0.7) \times 10^{65}$ & $(1.0 \pm 0.2) \times 10^{63} $ \\
(14)  & $F_{2-10} \ ({\rm erg \ cm^{-2} \ s^{-1}}$) & $7.14 \times 10^{-13}$ & $8.03 \times 10^{-12}$  \\
(15)  & $F_{10-50} \ ({\rm erg \ cm^{-2} \ s^{-1}}$) & $7.96 \times 10^{-12}$ & $3.76 \times 10^{-11}$  \\
(16) & $L_{2-10} \ ({\rm erg \ s^{-1}}$) & $1.76 \times 10^{43}$ &  $5.61 \times 10^{42}$ \\
     & $\chi^2$/dof   & 81.7/78 & 182.6/182 
\enddata
\tablecomments {
Errors correspond to $90\%$ confidence level for a single parameter.
Both $f_{\rm scat}$ and $R$ are normalized to the direct component
flux as measured with BAT.
\\
(1) The hydrogen column density of Galactic absorption by \citet{Kal05}. \\
(2) The line-of-sight hydrogen column density for the direct component. \\
(3) Normalization ratio of the direct component between the BAT and 
{\it Suzaku} spectra. \\ 
(4) The power-law photon index of the direct component. \\ 
(5) The fraction of the scattered component relative to the intrinsic power law.\\
(6) The center energy of the iron-K$\alpha$ emission line at the rest frame. \\
(7) The observed width of iron-K$\alpha$ line. This is fixed at 1 eV when we determine the continuum parameters. \\
(8) The observed equivalent width of the iron-K$\alpha$ line. \\
(9) The observed equivalent width of the iron-K$\alpha$ line with respect to the reflection component. \\
(10) The line-of-sight hydrogen column density for the reflection component. \\
(11) The relative strength of the reflection component to the direct one, defined as $R \equiv \Omega / 2 \pi$, where $\Omega$ is the solid angle of the reflector.\\
(12) The temperature of the {\bf apec} component. \\
(13) The emission measure of the {\bf apec} component. \\
(14) The observed {\it Suzaku} flux in the 2--10 keV band. \\
(15) The observed {\it Suzaku} flux in the 10--50 keV band. \\
(16) The 2--10 keV intrinsic luminosity obtained with {\it Suzaku},
corrected for the absorption.\\ 
}
\end{deluxetable*}

\subsubsection{IC 5063}\label{IC 5063}

Model~B with thin thermal emission ({\bf apec}) is adopted as the
best model for IC~5063 from the time averaged {\it Suzaku}
spectra. The thermal emission is required because 
it significantly reduces $\chi^2$  in  all 3 models (A, B, and C) and its addition produces acceptable fits.
Models A, B, and C give $\chi^2/{\rm dof}$ =
185.8/177, 175/175 and 175/173, respectively. Thus, the improvement of
the fit by adding a reflection component (Model B) is found to be
significant with an F-test probability of 0.007, whereas the introduction of 
an additional absorber for the transmitted component (Model C) is not
significant in introducing $\chi^2$. The results of simultaneous analysis of the {\it
Suzaku} and BAT spectra with Model~B + {\bf apec} are summarized
in Table~\ref{para}, and the best-fit model is plotted in
Figures~\ref{data} and \ref{model}. 

The best fit photon index of $\Gamma \approx 1.7$ from IC~5063
is typical of radio galaxies. The scattered fraction $f_{\rm
scat} \sim 0.9\%$ is larger than that of 3C~403. 
It is consistent with the {\it ASCA} result of $\sim$1\% \citep{Vig97}.
The narrow iron-K line detected at 6.4 keV is consistent with being
emitted from the torus. Its equivalent width with respect to the
reflected continuum ($1.5\pm0.2$ keV) is within a range of theoretical
expectation, thus supporting the validity of our spectral
model. Similar to the case of 3C~403, the reflection component is less
absorbed ($N_{\rm H}^{\rm refl} \approx 3\times 10^{22}$ \cmsq)
compared with the transmitted component (\nh $\approx 3\times 10^{23}$
\cmsq), and disfavoring the likely edge-on geometry. 
The reflection strength of $R \approx
0.6$ relative to the averaged BAT flux is close to the value obtained
from 3C~403. These results suggest that the torus geometry of the two
AGNs could be similar to each other. The obtained temperature and
luminosity of the {\bf apec} component are in the typical range of hot
ISM detected from elliptical galaxies \citep{Mat00} and hence supports
this interpretation of its origin, although we cannot rule out the
possibility that a part of this emission may come from photoioinized
plasma irradiated by the AGN, which should be regarded as an
accompanying component of the scattered emission. Nevertheless, since
the amplitude of the scattered fraction is mainly determined from 
the spectrum in 1--2 keV band, where no strong emission lines are present, 
its intensity is not strongly affected by the uncertainty in the nature of the line emitting component.

Since IC~5063 exhibits significant time variability during the {\it
Suzaku} observation, we check time variability of the spectra by
dividing the time region into two states with different flux levels,
epoch I (for 57.6 ks from the beginning) and epoch~II (after that).
We adopt the Model B + {\bf apec} model for each spectrum. Assuming
that the reflection component and thin thermal emission were not
variable on this short time scale, since they most probably originate from
the torus and host galaxy, respectively, we fix the spectral
parameters of these components (and its absorption) at the best-fit
values determined by the time-averaged {\it Suzaku} (and {\it
Swift}/BAT) spectra. We find that the shape of the transmitted
component did not show variability within statistical errors, 
$\Gamma = 1.7 \pm 0.1$ and $N_{\rm H} = (26 \pm 2) \times 10^{22}$ \cmsq
in epoch~I and $\Gamma = 1.8 \pm 0.1$ and $N_{\rm H} = (25 \pm 2) \times 
10^{22}$ \cmsq in epoch~II, and only the
normalization changed between the two epochs. Thus, averaging the {\it
Suzaku} spectra over the whole epoch is justified.

\begin{deluxetable*}{cccc}
\tabletypesize{\footnotesize}
\tablecaption{Best-Fit Parameters of {\it Suzaku} and {\it Swift}/BAT spectra with Torus Model\label{para_torus_model}}
\tablewidth{0pt} 
\tablehead{\colhead{} & \colhead{} & \colhead{3C 403} & \colhead{IC 5063} }
\startdata
(1)  & Model & Solar + apec$^{\rm a}$ + two power laws & Solar + apec$^{\rm a}$ \\
(2)  & \nhg ($10^{22}$ \cmsq) & 0.122$^{\rm b}$ & 0.061$^{\rm b}$ \\
(3)  & $N_{\rm H}^{\rm eq}$ ($10^{22}$ \cmsq) & $67 \pm 25$ & $25^{+10}_{-1}$ \\
(4)  & $\theta_{\rm oa}$ (degrees) & 50$^{\rm b}$ & 50$^{\rm b}$ \\
(5)  & $\theta_{\rm inc}$ (degrees) & $51^{+17}_{-0}$ & $65^{+18}_{-14}$\\
(6)  & Norm$_{\rm BAT}$ & $ 3.2 \pm 1.2$ & $ 1.04^{+0.16}_{-0.15}$ \\
(7)  & $\Gamma$  & $1.55^{+0.32}_{-0.05}$ & $1.82^{+0.08}_{-0.11} $ \\
(8)  & $f_{\rm scat, 0}$(\%)  & $0.3 (< 0.7)$ & $0.5 \pm 0.1$ \\
(9)  & $E_{\rm cen}$ (keV)   & 6.4$^{\rm b}$  & 6.4$^{\rm b}$ \\
(10)  & $\epsilon_{\rm Fe}$ & $1.2^{+0.7}_{-0.5}$ & $1.2^{+0.2}_{-0.3}$\\
     & $\chi^2$/dof   & 88.4/80 & 182.0/187
\enddata
\tablecomments {
Errors are $ 90 \% $ confidence level for a single parameter. 
$f_{\rm scat, 0}$ is normalized to the direct component
flux as measured with BAT. \\
(1) The model used in the fit. ``Solar'' means the torus model with Solar abundances. \\
(2) The hydrogen column density of Galactic absorption by \citet{Kal05}. \\
(3) The hydrogen column density of the torus along the equatorial direction. \\ 
(4) The half opening angle of the torus. \\ 
(5) The inclination angle of the torus. The lower limit is restricted to be 51$^\circ$ by the condition $\theta_{\rm inc} > \theta_{\rm oa} + 1$.\\
(6) Normalization ratio of the direct component between the BAT and {\it Suzaku} spectra. \\ 
(7) The power-law photon index. It is restricted in the range of 1.5--2.5 in the table model.\\
(8) The fraction of the scattered component relative to the intrinsic power law when the half opening angle of the torus is 45$^\circ$. \\
(9) The center energy of the iron-K emission line at the rest frame of the source redshift. \\
(10) The relative strength of the iron-K emission line to that predicted by the torus model. \\
$^{\rm a}$ The parameters of the {\bf apec} model are fixed at those in Table~3. \\
$^{\rm b}$ The parameters are fixed at this value.
}
\end{deluxetable*}

\subsection{Application of Torus Model}\label{torus_model}

We also fit the spectra generated by the Monte-Carlo model of \citet{Ike09}, where reflection
components from a three-dimensional,
axially symmetric torus around the nucleus emitting a cutoff power law
continuum is computed. The torus has
four parameters, the half opening angle ($\theta_{\rm oa}$),
inclination ($\theta_{\rm inc}$), total column density in the
equatorial plane ($N_{\rm H}^{\rm eq}$), and the ratio between the
outer and inner radii, which is fixed to be 100 (for definition, see
Figure~2 of \citet{Ike09}).

We analyze the data in the same manner as described in \citet{Egu11}, 
to which we refer the reader for details of the
spectral formula in the XSPEC terminology. 
The unabsorbed and absorbed reflection continua
and fluorescence iron-K line from the torus 
are included in the model as separate components.
The line-of-sight column density of the direct component is
automatically determined from the four geometrical parameters of the
torus; for absorbed AGNs $\theta_{\rm oa} < \theta_{\rm inc}$. The
observed scattered component is represented as
$f_{\rm scat} = f_{\rm scat,0}(1-\cos{\theta_{\rm oa}})/(1-\cos{45^\circ})$,
where $f_{\rm scat,0}$ reflects the averaged column density of the scattering gas. 
The reflection component from the accretion disk is also considered for
the direct component by assuming a solid angle of $\Omega=2\pi$ with
the same inclination as for the torus. Solar abundances are
assumed. We fit the {\it Suzaku} (XIS and HXD/PIN) and {\it Swift}/BAT
spectra simultaneously by limiting the energy band to 0.5--100 keV to
which the Ikeda model is applicable. Although the model uses a
cutoff energy of 360 keV instead of 200 keV, the effect is negligible
as we only use energies below 100 keV. Similar to the analysis in
Section~\ref{spec_model}, time variability of the flux 
of the direct component between the {\it
Suzaku} and {\it Swift}/BAT epochs is taken into account with an
assumption that the power law slope is constant. The {\bf apec}
component and the two power law components originating from the
radio lobes/wings and resolved jet components (for 3C~403) are
included in the spectral model, whose parameters are fixed at the
values found in Section~\ref{3C 403}.

We find that this ``torus model'' also provides a good fit to the
observed spectra of 3C~403 and IC~5063 with $\chi^2$/{\rm dof}=88/80
and 182/187, respectively. The torus opening angle is not well
determined in the model due to the limited photon statistics and we
fixed it at $\theta_{\rm oa}=50^\circ$ in our analysis consistent with
the optical observations of the ionized cone in IC~5063 \citep[$\sim
50^\circ$;][]{Col91}. The free parameters of the spectral model are
$\theta_{\rm inc}$, $N_{\rm H}^{\rm eq}$, $f_{\rm scat,0}$, the
normalization ratio between the iron-K emission line and reflected
continuum ($\epsilon_{\rm Fe}$), the photon index ($\Gamma$), and two
normalizations of the cutoff power law for the {\it Suzaku} and {\it
Swift}/BAT data. The best-fit parameters are summarized in
Table~\ref{para_torus_model}. We find $f_{\rm scat,0} = 0.3 (<
0.7)\%$ $(1.0^{+0.8}_{-0.5}\%)$ for 3C~403 and $f_{\rm scat,0} = 0.5
\pm 0.1\%$ $(0.6^{+0.1}_{-0.2}\%)$ for IC 5063 when normalized to the
BAT (or {\it Suzaku}) flux. These results on the scattering gas
will be discussed in Section~4. 
In both targets, the best-fit photon
index becomes slightly larger than in the analytical model,
although the difference is within the statistical errors. The
normalization ratio of the direct component between the BAT and {\it
Suzaku} spectra is consistent with that obtained from the analytic
model. As discussed in the previous subsections, the data indeed favor
an inclination angle that is only slightly larger than the torus
opening angle, particularly for 3C~403.

A similar numerical torus model by \citet{Mur09} (MYTORUS in XSPEC)
also gives a good fit for both targets, where a slightly different
torus geometry is assumed with a fixed opening angle of $60^\circ$.
By adopting it instead of the Ikeda model in the same way, we obtain
$\Gamma=1.7^{+0.1}_{-0.3}$, $\theta_{\rm inc} =75^{+10}_{-14}$ deg, 
and $N_{\rm H}^{\rm eq}=0.8^{+2.1}_{-0.3} \times 10^{24}$ \cmsq 
for 3C~403 with $\chi^2$/{\rm dof}=85.5/79, and
$\Gamma=1.9\pm0.1$, $\theta_{\rm inc} =63^{+9}_{-2}$ deg, and 
$N_{\rm H}^{\rm eq}=0.6^{+0.6}_{-0.2} \times 10^{24}$ \cmsq for 
IC~5063 with $\chi^2$/{\rm dof}=182.3/183.  
These parameters are consistent with those obtained by the Ikeda model 
within the statistical errors, except for $N_{\rm H}^{\rm eq}$ of 
IC 5063.

\section{Summary and Discussion}\label{discussion}

With {\it Suzaku}, we have for the first time obtained simultaneous
broad band X-ray spectra from two NLRGs, 3C 403 and IC 5063, with CCD
energy resolution below 10 keV. Combined with the {\it Swift}/BAT
spectra averaged for 58 months, we are able to best constrain the
spectral properties over the 0.5--200 keV band. The bolometric
luminosities of 3C 403 and IC 5063 are estimated as
$10^{45.1}$~erg~s$^{-1}$ and $10^{44.3}$~erg~s$^{-1}$ 
from the infrared and X-ray luminosities \citep{Vas10},
corresponding to the Eddington fractions of approximately 0.06,
using the black hole masses estimated by
\citet{Vas10}.
These Eddington fractions are in the range observed for radio galaxies
\citep[e.g.,][]{Eva06} and for Seyfert galaxies \citep{Win09} 
in the local universe.
Our targets thus significantly increase the NLRG sample with similar Eddington fractions studied with {\it Suzaku} and {\it Swift}/BAT in addition to NGC~612, which has $L_{\rm bol}/L_{\rm Edd}$ = 0.02 and $L_{\rm bol}$ = $10^{44.5}$~erg~s$^{-1}$ \citep{Vas10}.

While {\it Suzaku} provide us with unique spectroscopic data, the
integrated spectra from radio galaxies inevitably contain the
contribution from various components besides the nucleus, such as the
hot ISM, the radio lobes/wings, and the kpc-scale jets, due to the
limited spatial resolution. The thermal emission is important only 
in the softest energy band in the {\it Suzaku} data.
In the case of 3C~403, we are able to 
separate those from the radio lobes/wings and kpc-jets,
utilizing the {\it Chandra} results \citep{Kra05}. We find that the
relative fluxes of these components are 1.0\% and 0.7\% of the
unabsorbed nucleus emission at 5 keV, respectively, and thus make
only minor contribution in the hard X-ray spectra. Although we cannot
do the same analysis for IC~5063 and NGC~612 due to the lack of {\it
Chandra} data, we estimate that the inverse Compton component
originating from the radio lobes is 0.1\% and 1\% of the nucleus
emission at 5 keV, respectively, using the observed radio spectra from
the lobes \citep{Jon92} and a typical X-ray (inverse Compton) to radio
(synchrotron) flux ratio of the lobe emission \citep{Iso02}. One
should note, however, that emission from the ``nucleus'' may also contain
components from the pc-scale jets, which are seen in the VLBI
images \citep{Tar07,Oos00} but cannot be resolved with {\it Chandra}.  In
fact, \citet{Eva06} and \citet{Har09} suggest that the unabsorbed soft
X-ray components of radio galaxies is mainly attributable to the
unresolved jets, based on the correlation between the soft X-ray and
radio core luminosities at 5 GHz. In this context, the ``scattered
fraction'' we discuss in our paper is only an upper limit.

For 3C~403, we can compare our results of the scattered fraction and 
absorption column density of the transmitted component obtained 
through X-ray spectroscopy with those estimated by \citet{Mar05}
from the near-infrared (NIR) and optical imaging data.
\citet{Mar05} calculate the scattered fraction and extinction from the
``observed'' ratios of the optical (continuum or line) luminosity to the
NIR one, assuming the typical ``intrinsic'' luminosity ratios obtained for
radio loud AGNs.
Our result, $f_{\rm scat} = (0.4 \pm 0.3)$\%, is smaller than their
estimate, $0.88^{+0.32}_{-0.23}$\%, though within the errors. 
This possible discrepancy may be explained
if there is another unabsorbed component in the NIR/optical lights in
addition to the scattered one, such as thermal emission from hot dust
in the torus, and/or if there was time variability in the direct
NIR/optical fluxes, which were measured from short term
observations. The column density derived from the X-ray data is $\sim$10 times larger than that converted from the estimated
extinction, $A_{\rm V} = 21.9$ mag, assuming the Galactic standard
relation $A_{\rm V} = 5 \times 10^{-22} N_{\rm H}$ \citep{Boh78}. This
suggests much smaller dust-to-gas ratio in the torus than Galactic
gas, as reported for many other AGNs \citep[e.g.,][]{Mai01}.

Applying analytic spectral models that are commonly used to represent
the X-ray nuclear spectra of AGNs (i.e., a cutoff power law and
reflection component), we constrain the reflection strength in terms
of the solid angle of the reflector to be $R\equiv\Omega/2\pi \approx
0.6$ for both 3C~403 and IC~5063. A similar value ($R\approx0.5$ when
normalized to the BAT flux) is obtained for NGC~612 by \citet{Egu11},
while a weaker reflection ($R<0.4$ at 90\% confidence limit) is found
from another NLRG, 3C~33. These results suggest that a major fraction
of NLRGs show only a moderately strong reflection component 
in their X-ray spectra, like
many Seyfert 2 galaxies \citep{Dad07}, implying that their tori have a
sufficiently large solid angle to produce the reprocessed
emission. Thus, there is no evidence that the torus structure is
systematically different between radio galaxies and Seyfert galaxies,
although studies of a larger NLRG sample are required to investigate
this issue. The reason for the weak reflection in 3C~33 is unclear but
may be connected to its high X-ray luminosity ($L \sim 4 \times
10^{44}$~erg~s$^{-1}$ in the 2--100 keV band) compared with the other
NLRGs.

The application of the numerical torus model where a realistic
geometry is assumed gives useful insights into the torus structure, as
demonstrated in our previous studies \citep[e.g.,][]{Ike09,Awa09,Egu11}.
For our targets, we cannot constrain the torus opening angle due to
the limited photon statistics; in 3C~403 since values of $\theta_{\rm
oa}$ = $30^\circ$, $50^\circ$, or $70^\circ$ yields a very similar
$\chi^2$ value, and thus we cannot determine if the torus of 3C~403 is
close to that of ``new type'' AGNs or to ``classical type'' 
\citep{Ued07,Egu09} from our data alone. 
As for IC~5063 we have a
reasonable constraint on the torus geometry from the measurement of
the ionization cone, $\theta_{\rm oa}=50^\circ$. Together with the
fact that NGC~612 has $\theta_{\rm oa} > 58^\circ$ \citep{Egu11}, we
do not find any evidence for the presence of ``new type'' AGNs in
NLRGs studied with {\it Suzaku} so far, although we cannot rule out
its possibility for 3C~403.
We need a large sample with good quality broad band spectra to
verify whether or not the distribution of population is bimodal with
``new type'' and ''classical type'' as proposed by \citet{Egu09}, both
for radio quiet and loud AGNs.

A notable feature seen in 3C~403, IC~5063, and NGC~612,
is the small amount of scattering gas, parameterized as $f_{\rm
scat,0}$, reflecting the averaged column density of the scattering gas.
We find $f_{\rm scat,0}
< 0.7\%$ in these radio galaxies, which is systematically smaller than 
that obtained for Seyfert galaxies fitted with the same model, 
$> 2.0$\% for NGC~3081 \citep{Egu11} and 1.3$^{+0.3}_{-0.4}$\% for 
SWIFT~J0255.2--0011 \citep{Egu11b}, in which the scattered 
component is normalized to the flux as measured with BAT. As mentioned above,
one must take this value only as an upper limit considering the
possible contribution from the unresolved jets to the soft X-ray
component.  The $f_{\rm scat,0}$ value depends on the torus opening
angle assumed, but the above conclusion is unchanged unless
$\theta_{\rm oa}$ $\ll$ $30^\circ$ for 3C~403. Our result suggests
that there is intrinsically little ``normal'' gas around the nucleus
in radio galaxies. This result may be explained by the jet activity 
which can expel such gas.

We note that the obtained reflection strengths and scattered fractions
in 3C~403 and IC~5063 are normalized to the hard X-ray flux over 58
months obtained through the {\it Swift}/BAT monitoring, which should
give a good measure of the ``averaged'' activity of each source. In
fact, without the {\it Swift}/BAT data, we obtain an apparently very
strong reflection strength $R\sim 2$ from 3C~403. We demonstrate that
such an unphysically large $R$ value can indeed be accounted for by
time variability, as discussed in e.g., \citet{Ued07}; the flux of the
direct component was much fainter than the average flux in the past
that is responsible for the reprocessed emission from large scale
($>$pc) surroundings. Thus, the combination of the simultaneously
obtained {\it Suzaku} spectra and the long averaged {\it Swift}/BAT
is necessary to estimate a ``true'' reflection strength and
scattered fraction that are free from the effects of short-term 
time variability.

\acknowledgments

We thank Dr. Wayne Baumgartner for sending us the {\it Swift}/BAT
spectra of 3C~403 and IC~5063 when the web server was down. This work
was partly supported by the Grant-in-Aid for Scientific Research
20540230 (YU) and 20740109 (YT), and by the grant-in-aid for the
Global COE Program ``The Next Generation of Physics, Spun from
Universality and Emergence'' from the Ministry of Education, Culture,
Sports, Science and Technology (MEXT) of Japan.


\begin{thebibliography}{}

\bibitem[Anders \& Grevesse(1989)]{And89} 
  Anders, E., \& Grevesse, N.\ 1989, \gca, 53, 197

\bibitem[Awaki et al.(2009)]{Awa09} 
  Awaki, H., Terashima, Y., Higaki, Y., \& Fukazawa, Y.\ 2009, \pasj, 61, 317


\bibitem[Baumgartner et al.(2011)]{Bau11}
  Baumgartner, W., et al.\ 2011, \apjs, submitted

\bibitem[Bohlin et al.(1978)]{Boh78}
  Bohlin, R. C., Savage, B. D. \& Drake, J. F.\ 1978, \apj, 224. 132

\bibitem[Colina et al.(1991)]{Col91}
  Colina, L., Sparks, W. B., \& Macchetto, F.\ 1991, \apj, 370, 102

\bibitem[Dadina(2007)]{Dad07}
  Dadina, M.\ 2007, \aap, 461, 1209

\bibitem[Davies et al.(2007)]{Dav07}
  Davies, R. I., M\"{u}ller S\'{a}nchez, F., Genzel, R., Tacconi, L. J., Hicks, E. K. S., Friedrich, S., \& Sternberg, S.\ 2007, \apj, 671, 1388

\bibitem[Dennett-Thorpe et al.(2002)]{Den02}
  Dennett-Thorpe, J., Scheuer, P. A. G., Laing, R. A., Bridle, A. H.,
  Pooley, G. G., \& Reich, W.\ 2002, \mnras, 330, 609

\bibitem[Di Matteo et al.(2005)]{DiM05}
  Di Matteo, T., Springel, V., \& Hernquist, L.\ 2005, \nat, 433, 604

\bibitem[Deluit \& Courvoisier(2003)]{Del03}
  Deluit, S., \& Courvoisier, T. J. L.\ 2003, \aap, 339, 77

\bibitem[Eguchi et al.(2009)]{Egu09} 
  Eguchi, S., Ueda, Y., Terashima, Y., Mushotzky, R., \& Tueller, J.\ 2009, \apj, 696, 1657

\bibitem[Eguchi et al.(2011)]{Egu11} 
  Eguchi, S., Ueda, Y., Awaki, H., Aird, J., Terashima, Y., \& Mushotzky, R.\ 2011, \apj, 729, 31

\bibitem[Eguchi(2011)]{Egu11b}
  Eguchi, S.\ 2011, Doctoral Thesis, Kyoto University

\bibitem[Elitzur \& Shlosman(2006)]{Eli06}
  Elitzur, M., \& Shlosman, I.\ 2006, \apj, 648, 101


\bibitem[Evans et al.(2006)]{Eva06} 
  Evans, D. A., Worrall, D. M., Hardcastle, M. J., Kraft, R. P., \& Birkinshaw, M.\ 2006, \apj, 642, 96



\bibitem[Fukazawa et al.(2009)]{Fuk09}
  Fukazawa, Y., et al.\ 2009, \pasj, 61, 17


\bibitem[Grandi et al.(2006)]{Gra06}
  Grandi, P., Malaguti, G., \& Fiocchi, M.\ 2006, \apj, 642, 113

\bibitem[Gruber et al.(1999)]{Gru99} 
  Gruber, D. E., Matteson, J. L., Peterson, L. E., \& Jung, G. V.\ 1999, \apj, 520, 124

\bibitem[Hardcastle et al.(2009)]{Har09} 
  Hardcastle, M. J., Evans, D. A., \& Croston, J. H.\ 2009, \mnras, 396, 1929

\bibitem[Ikeda et al.(2009)]{Ike09} 
  Ikeda, S., Awaki, H., \& Terashima, Y.\ 2009, \apj, 692, 608

\bibitem[Imanishi \& Wada(2004)]{Ima04}
  Imanishi, M., \& Wada, K.\ 2004, \apj, 617, 214

\bibitem[Ishisaki et al.(2007)]{Ish07} 
  Ishisaki, Y., et al.\ 2007, \pasj, 59, 113


\bibitem[Isobe(2002)]{Iso02}
Isobe, N.\ 2002, Doctoral Thesis, University of Tokyo

\bibitem[Jones \& McAdam(1992)]{Jon92}
  Jones, P. A., \& McAdam, W. B.\ 1992, \apjs, 80, 137

\bibitem[Kalberla et al.(2005)]{Kal05}
  Kalberla, P. M. W., Burton, W. B., Hartmann, D., Arnal, E. M., Bajaja, E., Morras, R., P\"{o}ppel, W. G. L.\ 2005, \aap, 440, 775 


\bibitem[Komatsu et al.(2009)]{Kom09} 
  Komatsu, E., et al.\ 2009, \apj, 180, 330

\bibitem[Koyama et al.(1992)]{Koy92} 
  Koyama, K., Awaki, H., Iwasawa, K., \& Ward M. J.\ 1992, \apj, 399, 129


\bibitem[Kraft et al.(2005)]{Kra05} 
  Kraft, R. P., Hardcastle, M. J., Worrall, D. M., \& Murray, S. S.\ 2005, \apj, 622, 149


\bibitem[Leahy \& Williams(1984)]{Lea84} 
  Leahy, J. P., \& Williams, A. G.\ 1984, \mnras, 210, 929

\bibitem[Maeda et al.(2008)]{Mae08} 
  Maeda, Y., et al.\ 2008, JX-ISAS-SUZAKU-MEMO-2008-06

\bibitem[Magdziarz \& Zdziarski(1995)]{Mag95}
  Magdziarz, P., \& Zdziarski, A. A.\ 1995, \mnras, 273, 837

\bibitem[Magorrian et al.(1998)]{Mag98}
  Magorrian, J., et al.\ 1998, \aj, 115, 2285

\bibitem[Maiolino et al.(2001)]{Mai01}
  Maiolino, R., Marconi, A., Salvati, M., Risaliti, G., 
  Severgnini, P., Oliva, E., La Franca, F., \& Vanzi, L.\ 2001, 
  \aap, 365, 28

\bibitem[Marchesini et al.(2005)]{Mar05}
  Marchesini, D., Capetti, A., \& Celotti, A.\ 2005, \aap, 841, 854

\bibitem[Marconi \& Hunt(2003)]{Mar03}
  Marconi, A., \& Hunt, L. K.\ 2003, \apj, 589, 21


\bibitem[Matsushita et al.(2000)]{Mat00}
  Matsushita, K., Ohashi, T., \& Makishima, K.\ 2000, \pasj, 52, 685

\bibitem[Mitsuda et al.(2007)]{Mit07} 
  Mitsuda, K., et al.\ 2007, \pasj, 59, S1

\bibitem[Murphy \& Yaqoob(2009)]{Mur09}
  Murphy, K. D., \& Yaqoob, T.\ 2009, \mnras, 397, 1549

\bibitem[Oosterloo et al.(2000)]{Oos00}
  Oosterloo, T. A., Morganti, R., Tzioumis, A., Reynolds, J., King, E.,
  McCulloch, P., \& Tsvetanov, Z.\ 2000, \aj, 119, 2085






\bibitem[Tarchi et al.(2007)]{Tar07}
  Tarchi, A., Brunthaler, A., Henkel, C., Menten, K. M. Braatz, J., 
   \& Wei\ss, A.\ 2007, \aap, 475, 497


\bibitem[Tueller et al.(2008)]{Tue08}
  Tueller, J., et al.\ 2008, \apj, 681, 113



\bibitem[Ueda et al.(2007)]{Ued07}

  Ueda, Y., et al.\ 2007, \apjl, 664, L79

\bibitem[Vasudevan et al.(2010)]{Vas10}
  Vasudevan, R. V., Fabian, A. C., Gandhi, P., Winter, L. M., \& Mushotzky, R. F.\ 2010, \mnras, 402, 1081

\bibitem[Vasudevan \& Fabian(2009)]{Vas09}
  Vasudevan, R. V., \& Fabian, A. C.\ 2009, \mnras, 392, 1124

\bibitem[Vignali et al.(1997)]{Vig97}
  Vignali, C., Comastri, A., Cappi, M., \& Palumbo, G. G. C.\ 1997, 
  \memsai, 68, 139

\bibitem[Winter et al.(2009)]{Win09}
  Winter, L. M., Mushotzky, R. F., Terashima, Y., \& Ueda, Y.\ 2009,
  \apj, 701, 1644

\end{thebibliography}
\end{document}